\documentclass{article}
\usepackage{hiph-art}
\newcommand{\NP}[1]{ Nucl.\ Phys.\ {#1}}

\newcommand{\PR}[1]{Phys.\ Rev.\ {#1}}

\newcommand{\La}{{\cal L}}

\newcommand{\be}{\begin{equation}}
\newcommand{\ee}{\end{equation}}
\newcommand{\ba}{\begin{eqnarray}}
\newcommand{\ea}{\end{eqnarray}}

\volnumber{19} \issuenumber{1} \edyear{2004}                             
\frompage{000} \topage{000}                                              
\recrevdate{1 January 2004}                                              
\def\rightmark{Chiral dynamics of hadrons in nuclei}
\def\leftmark{A. One and A. Two}
 
\title{Chiral dynamics of hadrons in nuclei} 
\authors{ 
{E. Oset$^1$, D. Cabrera$^1$, V.K. Magas$^1$, L. Roca$^1$, M.J. Vicente Vacas$^1$,
A. Ramos$^{2}$, T. Inoue$^{3}$, C. Garcia Recio$^{4}$ and J. Nieves$^{4}$. %
\index{One, A.} 
\index{Two, A.} 
}\\[2.812mm]
{\normalsize
\hspace*{-8pt}$^1$ Departamento de F\'{\i}sica Te\'orica and IFIC,
Centro Mixto Universidad de Valencia-CSIC,
Institutos de Investigaci\'on de Paterna, Apdo. correos 2085,\\
46071, Valencia, Spain.\\[0.2ex] 
\hspace*{-8pt}$^2$ Departament d'Estructura i Constituents de la Materia\\ 
Universitat de Barcelona, 08028 Barcelona, Spain.\\
\hspace*{-8pt}$^3$ Institut fur Theoretische Physik, Universitaet Tuebingen\\
Tuebingen, Germany.\\
\hspace*{-8pt}$^4$ Departamento de Fisica Moderna, Universidad de Granada\\
Granada, Spain.
}}
 
\abstract{In this talk I report on selected topics of hadron modification in the
nuclear medium using the chiral unitary approach to describe 
the dynamics of the
problems. I shall mention how antikaons, $\eta$, and $\phi$ are modified in the
medium and will report upon different  experiments done or planned to 
measure the $\phi$ width 
in the medium.}

\PACS{see: http://www.aip.org/pacs/ }
 
\makeindex
\begin{document}
 
\maketitle

\section{Introduction}\label{intro}
 Nowadays it is commonly accepted that QCD is the theory of the strong
interactions, with the quarks as building blocks for baryons and mesons, and
the gluons as the mediators of the interaction. However, at low energies typical
of the nuclear phenomena, perturbative calculations with the QCD Lagrangian are
not possible and one has to resort to other techniques to use the information of
the QCD Lagrangian. One of the most fruitful approaches has been the use of
chiral perturbation theory, $\chi PT$ \cite{xpt}. The theory introduces effective
Lagrangians which involve only observable particles, mesons and baryons,
respects the basic symmetries of the original QCD Lagrangian, particularly
chiral symmetry, and organizes these effective Lagrangians according to the
number of derivatives of the meson and baryon fields.  
 
\section{Baryon meson interaction}\label{baryon}
  The interaction of the octet of stable baryons with the octet of pseudoscalar
  mesons is given to lowest order by the Lagrangian \cite{ulf,ecker}
\begin{eqnarray}
\label{BaryonL}
\La_1=<\bar{B}i\gamma^\mu \bigtriangledown_\mu B >-M_B
<\bar{B}B>&+&\frac{1}{2}D<\bar{B}\gamma^\mu \gamma_5
\left\{u_\mu,B \right\}>\\
&+&\frac{1}{2}F<\bar{B}\gamma^\mu \gamma_5 [u_\mu,B] >\nonumber
\label{eq:lowest}
\end{eqnarray}  
  with the usual definitions for the 
B and $\Phi$  SU(3) matrices, the covariant derivative $\bigtriangledown_\mu$
and the SU(3) matrix $u_\mu$ \cite{ulf,ecker}.

 \section{Unitarized chiral perturbation theory: N/D or dispersion relation
method}

  One can find a systematic and easily comprehensible derivation 
 of the  ideas of the N/D method applied for the first time to the meson baryon system in
 \cite{Oller:2000fj}, which we reproduce here below and which follows closely
 the similar developments used before in the meson meson interaction \cite{nsd}.
 One defines the transition $T-$matrix as $T_{i,j}$ between the coupled channels which couple to
 certain quantum numbers. For instance in the case of  $\bar{K} N$ scattering studied in
 \cite{Oller:2000fj} the channels with zero charge are $K^- p$, $\bar{K^0} n$, $\pi^0 \Sigma^0$,$\pi^+
 \Sigma^-$, $\pi^- \Sigma^+$, $\pi^0 \Lambda$, $\eta \Lambda$, $\eta \Sigma^0$, 
 $K^+ \Xi^-$, $K^0 \Xi^0$.
 Unitarity in coupled channels is written as
 
\begin{equation} 
Im T_{i,j} = T_{i,l} \rho_l T^*_{l,j}
\end{equation}
where $\rho_i \equiv 2M_l q_i/(8\pi W)$, with $q_i$  the modulus of the c.m. 
three--momentum, and the subscripts $i$ and $j$ refer to the physical channels. 
 This equation is most efficiently written in terms of the inverse amplitude as
\begin{equation}
\label{uni}
\hbox{Im}~T^{-1}(W)_{ij}=-\rho(W)_i \delta_{ij}~,
\end{equation}
The unitarity relation in Eq. (\ref{uni}) gives rise to a cut in the
$T$--matrix of partial wave amplitudes, which is usually called the unitarity or right--hand 
cut. Hence one can write down a dispersion relation for $T^{-1}(W)$ 
\begin{equation}
\label{dis}
T^{-1}(W)_{ij}=-\delta_{ij}\left\{\widetilde{a}_i(s_0)+ 
\frac{s-s_0}{\pi}\int_{s_{i}}^\infty ds' 
\frac{\rho(s')_i}{(s'-s)(s'-s_0)}\right\}+{\mathcal{T}}^{-1}(W)_{ij} ~,
\end{equation}
where $s_i$ is the value of the $s$ variable at the threshold of channel $i$ and 
${\mathcal{T}}^{-1}(W)_{ij}$ indicates other contributions coming from local and 
pole terms, as well as crossed channel dynamics but {\it without} 
right--hand cut. These extra terms
are taken directly from $\chi PT$ 
after requiring the {\em matching} of the general result to the $\chi PT$ expressions. 
Notice also that 
\begin{equation}
\label{g}
g(s)_i=\widetilde{a}_i(s_0)+ \frac{s-s_0}{\pi}\int_{s_{i}}^\infty ds' 
\frac{\rho(s')_i}{(s'-s)(s'-s_0)}
\end{equation}
is the familiar scalar loop integral.

One can further simplify the notation by employing a matrix formalism. 
Introducing the 
matrices $g(s)={\rm diag}~(g(s)_i)$, $T$ and ${\mathcal{T}}$, the latter defined in 
terms 
of the matrix elements $T_{ij}$ and ${\mathcal{T}}_{ij}$, the $T$-matrix can be written as:
\begin{equation}
\label{t}
T(W)=\left[I-{\mathcal{T}}(W)\cdot g(s) \right]^{-1}\cdot {\mathcal{T}}(W)~.
\end{equation}
which can be recast in a more familiar form as 
 \begin{equation}
\label{ta}
T(W)={\mathcal{T}}(W)+{\mathcal{T}}(W) g(s) T(W)
\end{equation}
Now imagine one is taking the lowest order chiral amplitude for the kernel as done in
\cite{Oller:2000fj}. Then the former equation is nothing but the Bethe Salpeter equation with the
kernel taken from the lowest order Lagrangian and  factorized  on  shell, the same
approach followed in \cite{kaon}, where different arguments were used to justify the on shell
factorization of the kernel.

 \section{Meson baryon scattering}
The low-energy $K^-N$ scattering and transition to coupled channels is one of
the cases of successful application of chiral dynamics in the baryon
sector.  We rewrite Eq. (7) in the more familiar form 
\begin{equation}
T = V + V \, G \, T
\label{eq:bs2}
\end{equation}
with $G$ the diagonal matrix given by the loop function of a meson and a baryon
propagators.

The analytical expression for $G_l$ can be obtained from \cite{kaon} using a
 cut off and from \cite{Oller:2000fj} using dimensional regularization. 

\section{Strangeness $S= -1$ sector}

We take the $K^- p$ state and all those that couple to it within the chiral
scheme mentioned above.  Hence we have a problem with ten coupled channels.
The coupled set of Bethe Salpeter equations 
were solved in \cite{kaon} using a cut off momentum of 630
MeV in all channels. Changes in the cut off can be accommodated in terms
 of changes in $\mu$, the regularization scale in the dimensional
 regularization formula for  $G_l$, or in the subtraction constant
$a_l$. In
 order to obtain the same results as in \cite{kaon} at low energies, we set
 $\mu$ equal to the cut off momentum of 630 MeV (in all channels) and then
find the values of the
 subtraction constants $a_l$ such as to have $G_l$ with the same value
with the
 dimensional regularization formula  and the cut
off formula at the $\bar{K} N$ threshold. 
For the purpose of this talk let us recall that in \cite{kaon} we 
obtain the $\Lambda(1405)$ resonance 
obtained from the $\pi \Sigma$ spectrum and the cross sections for 
$K¯p $ to different channels, some of which are shown in fig. 1. 
Other resonances are also found but I shall not mention them in this talk.

\begin{figure}[htb]
\vspace*{-0.5 cm}
                 \insertplot{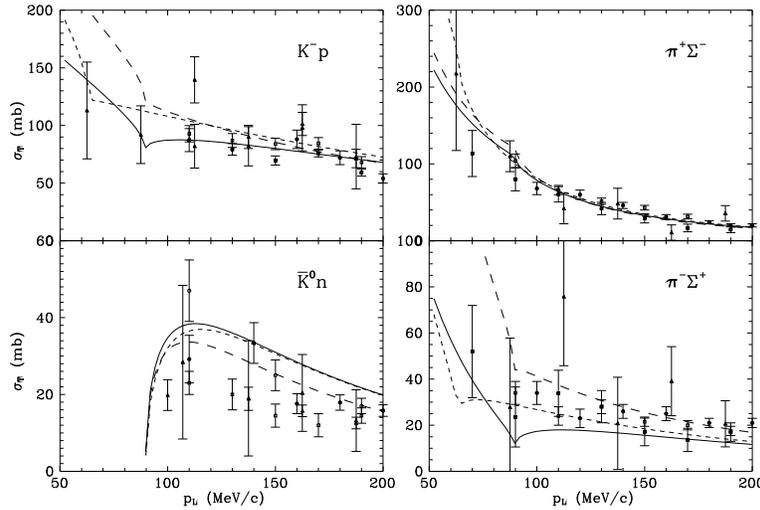}
\vspace*{-1.0 cm}
\caption{
$K^-p$ scattering cross sections as functions of the $K^-$
momentum in the lab frame: with the full basis of physical states
(solid line), omitting the $\eta$ channels (long-dashed line) and
with the isospin-basis (short-dashed line). Taken from
Ref.~\protect\cite{kaon}.
\label{fig:kncross}}
\end{figure}

\section{Strangeness $S= 0$ sector}

  The strangeness $S= 0$ channel was also investigated using the Lippmann
Schwinger equation and coupled channels in \cite{Kaiser:1995cy}. 
The $N^*(1535)$ resonance was also generated dynamically within this 
approach. Subsequently work was done in this sector in \cite{Nacher:1999vg},
and \cite{Nieves:2001wt} where the $N^*(1535)$ 
resonance was also generated. 
In \cite{Inoue:2001ip} the work along these lines was continued and improved
 by introducing the $\pi N \to \pi NN$ channels, which proved essential in
 reproducing the isospin 3/2 part of the $\pi N$ amplitude, including the
 reproduction of the $N^*(1650)$ resonance.

  For total zero charge one has six channels in this case,
 $\pi^- p$, $\pi^0 n$, $\eta n$, $K^+ \Sigma^-$,
 $K^0 \Sigma^0$, and $K^0 \Lambda$.   
 
 Details on the issues discussed in this
 paper can be seen in the review paper \cite{review}.
 
 \section{ $\bar{K}$ in nuclei}
 Next we address the properties of the $\bar{K}$ in the 
nuclear medium which have been studied in  \cite{knuc}. The work is based on
the elementary $\bar{K} N$ interaction which has been discussed above,
using a coupled channel unitary approach with chiral Lagrangians.

The coupled channel formalism requires to evaluate the transition
amplitudes between the different channels that can be built from
the meson and baryon octets. For $K^- p$ scattering there are 10 such
channels, namely $K^-p$, $\bar{K}^0 n$, $\pi^0
\Lambda$, $\pi^0 \Sigma^0$,
$\pi^+ \Sigma^-$, $\pi^- \Sigma^+$, $\eta \Lambda$, $\eta
\Sigma^0$,
$K^+ \Xi^-$ and $K^0 \Xi^0$. In the case of $K^- n$ scattering
the coupled channels are: $K^-n$, $\pi^0\Sigma^-$,
 $\pi^- \Sigma^0$, $\pi^- \Lambda$, $\eta
\Sigma^-$ and
$K^0 \Xi^-$.

 In order to evaluate
the $\bar{K}$ selfenergy in the medium, one needs first to include the medium 
modifications 
in the $\bar{K} N$ amplitude, $T_{\rm
eff}^{\alpha}$ ($\alpha={\bar K}p,{\bar K}n$), and then perform the
integral over the nucleons in the Fermi sea: 

\begin{equation}
\Pi^s_{\bar{K}}(q^0,{\vec q},\rho)=2\int \frac{d^3p}{(2\pi)^3}
n(\vec{p}) \left[ T_{\rm eff}^{\bar{K}
p}(P^0,\vec{P},\rho) +
T_{\rm eff}^{\bar{K} n}(P^0,\vec{P},\rho) \right] \ ,
\label{eq:selfka}
\end{equation}

The values
$(q^0,\vec{q}\,)$ stand now for the energy and momentum of the
$\bar{K}$ in the lab frame, $P^0=q^0+\varepsilon_N(\vec{p}\,)$,
$\vec{P}=\vec{q}+\vec{p}$ and $\rho$ is the nuclear matter density.

We also include a p-wave contribution to the ${\bar K}$ 
self-energy coming from the coupling of the ${\bar K}$ meson to
hyperon-nucleon hole ($YN^{-1}$) excitations,
with $Y=\Lambda,\Sigma,\Sigma^*(1385)$. The vertices $MBB^\prime$ are 
easily derived from
the $D$ and $F$ terms of Eq.~(1). The explicit expressions can be seen in
 \cite{knuc}. 
 
 At this point it is interesting to recall three different approaches to the
 question of the $\bar{K}$ selfenergy in the nuclear medium. The first
 interesting realization was the one in \cite{koch94,wkw96,waas97}, 
 where the Pauli blocking in the intermediate nucleon states 
 induced a shift of the $\Lambda(1405)$ resonance to higher
 energies and a subsequent attractive $\bar{K}$ selfenergy. The work of
 \cite{lutz} introduced a novel an interesting aspect, the selfconsistency.
 Pauli blocking required a higher energy to produce the resonance, but having a
 smaller kaon mass led to an opposite effect, and as a consequence the
 position of the resonance was brought back to the free position. Yet, a 
 moderate
 attraction on the kaons still resulted, but weaker than anticipated from the
 former work.  The work of \cite{knuc} introduces some novelties. It
 incorporates the selfconsistent treatment of the kaons done in \cite{lutz} 
 and in addition it also includes the selfenergy of the pions, which are let to
 excite ph and $\Delta h$ components. It also includes the  mean field
 potentials of the baryons.  
The obvious consequence of the work of \cite{knuc}
is that the spectral function of the kaons  gets much wider than in the two
former approaches because one is including new decay channels for the
$\bar{K}$ in nuclei.

  In the work of \cite{zaki} the kaon selfenergy discussed above has been 
  used for the case of kaonic atoms, where there are
abundant data to test the theoretical predictions. One uses the Klein 
Gordon equation and obtains two families of states. One
family corresponds to the atomic states, some of which are those already  
measured, and 
which have  energies around or below 1 MeV and widths
of about a few hundred KeV or smaller. The other family corresponds to 
states which are nuclear deeply bound states, with
energies of 10 or more MeV and widths around 100 MeV. 

With the $K^-$ many body decay channels included in our approach, the resulting
widths of the deeply bound $K^-$ states (never bound by more than 50 MeV) are
very large (of the order of 100 MeV) and, hence, there is no room for narrow
deeply bound $K^-$ states which appear in some oversimplified theoretical
approaches.  A recent phenomenological work \cite{Mares:2004pf} considering the
$K^- NN \to \Lambda N \Sigma N$ nuclear kaon absorption channels, which are also
incorporated in \cite{knuc}, also reaches the conclusion that in the unlikely
case that there would be deeply bound kaonic atoms they should have necessarily
a large width.

\section{ $\phi$ decay in nuclei}

Let us say a few words about the $\phi$ decay in nuclei. The work 
reported here \cite{phi} follows closely the lines of
\cite{klingl,norbert}, however, it uses the updated $\bar{K}$ 
selfenergies of \cite{knuc}.  In the present
case the $\phi$ decays primarily in $K\bar{K}$, but these kaons can 
now interact with the medium as discussed previously.    For
the selfenergy of the $K$, since the $KN$ interaction is not too strong 
and there are no resonances, the $t\rho$ approximation is
sufficient.
\begin{figure}[htb]
 \begin{center}
\insertplot{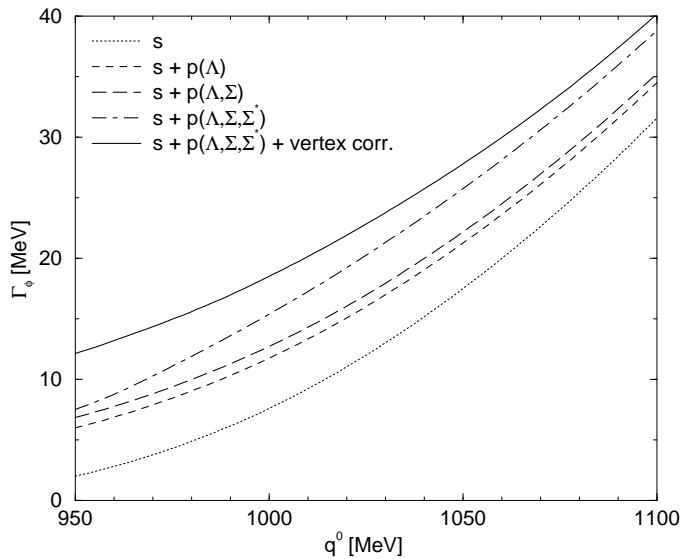}
 \caption{$\phi$ width at $\rho=\rho_0$.}
 \end{center}
\end{figure}
In fig. 2 we show the results for the $\phi$ width at $\rho=\rho_0$ 
as a function of the mass of the $\phi$, separating the
contribution from the different channels. What we observe is that the 
consideration of the s-wave $\bar{K}$-selfenergy is
responsible for a sizable increase of the width in the medium, but 
the p-wave is also relevant, particularly the $\Lambda h$
excitation and the $\Sigma^*h$ excitation. It is also interesting to 
note that the vertex corrections \cite{beng}  (Yh loops attached
to the $\phi$ decay vertex)  are now present and do not cancel off 
shell contributions like in the case of the scalar mesons. Their
contribution is also shown in the figure and has about the same 
strength as the other p-wave contributions.  The total width of
the $\phi$ that we obtain is about 22 MeV at $\rho=\rho_0$, about 
a factor two smaller than the one obtained in
\cite{klingl,norbert}. A recent evaluation of the $\phi$ selfenergy in the
medium \cite{dani} similar to that of \cite{phi}, in which also the real part is evaluated,
leads to a width about 20 percent larger than that of \cite{phi}. The important 
message from all these works is, however,  the 
nearly one order of magnitude increase of the width with respect to
the free one.

\section{$\eta$ selfenergy and eta bound in nuclei}
 The method of \cite{knuc} has also been used recently to determine the $\eta$ selfenergy
   in the nuclear medium \cite{Inoue:2002xw}.  One obtains a potential at threshold of the order 
   of (-54 -i29) MeV at
   normal nuclear matter, but it also has a strong energy dependence due to the proximity of
   the $N^*(1535)$ resonance and its appreciable modification in the nuclear
   medium. 
   
 To compute the $\eta-$nucleus bound states, we solve the Klein-Gordon 
 equation (KGE) with the $\eta-$selfenergy, 
 $\Pi_\eta(k^0,r) \equiv \Pi_\eta(k^0,\vec 0,\rho(r))$, obtained using the local
 density approximation.
 We have then:
 \begin{equation}
   \left[ -\vec\nabla^2 + \mu^2 +  \Pi_\eta(\mbox{Re}[E],r) \right] \Psi 
  = E^2 \Psi
 \end{equation}
 where $\mu$ is the $\eta-$nucleus reduced mass, the real part of $E$  
 is the total meson energy, including its mass, and the imaginary part
 of $E$, with opposite sign, is the half-width $\Gamma/2$ of the state. 
 The binding energy $B<0$ is defined as $B = \mbox{Re}[E]-m_{\eta}$.

 The results  from \cite{Garcia-Recio:2002cu} are shown in Table 1  for the 
 energy dependent potential.  On the
 other hand we see that the half widths of the states are  
 large, larger in fact
 than the binding energies or the separation energies between neighboring 
 states. 
 
\begin{table}[t]
\begin{center}
\caption{ (B,$-\Gamma/2$) for $\eta-$nucleus bound states 
calculated with the energy dependent potential. Units in MeV}
\label{tbl:statedep}
\footnotesize
\begin{tabular}{c|cccccc}
\hline
\hline 
  &   $^{12}$C  & $^{24}$Mg    &  $^{27}$Al    &  $^{28}$Si    &   $^{40}$Ca   &  $^{208}$Pb    \\
\hline 
1s&($-$9.71,$-$17.5)&($-$12.57,$-$16.7)&($-$16.65,$-$17.98)&($-$16.78,$-$17.93)&($-$17.88,$-$17.19)&($-$21.25,$-$15.88) \\
1p&             &              &( $-$2.90,$-$20.47)&( $-$3.32,$-$20.35)&( $-$7.04,$-$19.30)&($-$17.19,$-$16.58) \\
1d&             &              &               &               &               &($-$12.29,$-$17.74) \\
2s&             &              &               &               &
&($-$10.43,$-$17.99) \\
1f&             &              &               &               &               &( $-$6.64,$-$19.59) \\
2p&             &              &               &               &               &( $-$3.79,$-$19.99) \\
1g&             &              &               &               &               &( $-$0.33,$-$22.45) \\
\hline
\hline
\end{tabular}
\normalsize
\end{center}
\end{table}

   With the results obtained here it looks like the chances to see distinct
 peaks corresponding to $\eta$ bound states are not too big. 
 
   On the other hand one can look at the results with a more optimistic view
 if one simply takes into account that experiments searching for these states
 might not see them as peaks, but they should see some clear
 strength below threshold in the $\eta$ production experiments.
 The range by which this strength would go into the bound region
 would measure the combination of half width and binding energy. 
 Even if this is less information than the values of the energy and
 width of the states, it is by all means a relevant information to gain some
 knowledge on the $\eta$ nucleus optical potential.  
 
\section{Experiments to determine the $\phi$ width in the medium}

Recently there has been an experiment \cite{ishi} designed to determine the
width of the $\phi$ in nuclei.  Unlike other experiments proposed which aim at
determining the width from the $K^+ K^-$ invariant mass distribution and which
look extremely difficult \cite{Oset:2000na,muhlich}, the experiment done in
Spring8/Osaka uses a different philosophy, since it looks at the A dependence of
the $\phi$ photoproduction cross section. The idea is that the $\phi$ gets absorbed
in the medium with a probability per unit length equal to 
\begin{equation}
-Im \Pi /q
\end{equation}
where $\Pi$ is the $\phi$ selfenergy in the medium ($\Gamma=-Im \Pi /
\omega_{\phi}$). The bigger the nucleus the more $\phi$ get absorbed and there
is a net diversion from the A proportionality expected from a photonuclear
reaction.
 The method works and one obtains a $\phi$ width in the medium which is given in
 \cite{ishi} in terms of a modified $\phi N$ cross  section in the nucleus
 sizeably larger than the free one.  Prior to these experimental results there is
 a theoretical calculation in \cite{luis} adapted to the set up of the
 experiment of \cite{ishi}, based on the results for the $\phi $ selfenergy
 reported above.  The results agree only qualitatively with the experimental
 ones, the latter ones indicating that the $\phi$ selfenergy in the medium could
 be even  larger than the calculated one. However, although it has been fairly
 taken into account in the experimental analysis, there is an inconvenient in the
 reaction of \cite{ishi}, since there is a certain contamination of coherent
 $\phi$ production which blurs the interpretation of the data.
 
   In order to use the same idea of the A dependence and get rid of the coherent
$\phi$ production, a new reaction has been suggested in \cite{magas} which could
be implemented in  a facility like COSY. The idea is to measure the $\phi$
production cross section in different nuclei through the reaction
\begin{equation}
p A \to  \phi X
\end{equation}
The calculation is done assuming one step production, and two step production
with a nucleon or $\Delta$ in the intermediate states,  allowing for the loss of
$\phi$ flux as the $\phi$ is absorbed in its way out of the nucleus. Predictions
for the cross sections normalized to the one of $^{12}C$ are shown in fig. 3
where it is shown that with a precision of 10 percent in the experimental 
ratios one could
disentangle between the different curves in the figure and easily determine if
the width is one time, two times etc, the width determined in ref.
\cite{phi,dani} which is about 27 MeV for normal nuclear matter density. An
experiment of this type can be easily performed in the COSY facility, where
hopefully it will be done in the near future.

\begin{figure}[htb]
\begin{center}
\insertplot{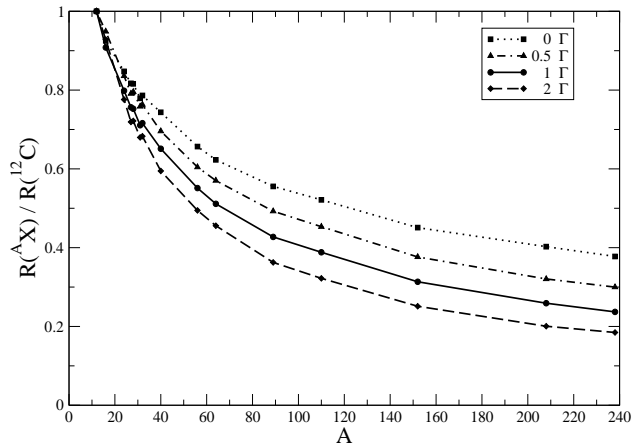}
\caption{Ratio of the nuclear cross section normalized to
$^{12}C$ for $T_p=2.83\textrm{ GeV}$ multiplying the $\phi$ width
in the medium, $\Gamma$ around 27 MeV, by different factors.} 
\end{center}
\end{figure}

\section*{Acknowledgments}
D.C. and  L.R. acknowledge support from the 
Ministerio de Educaci\'on y Ciencia.
This work is partly supported by the Spanish CSIC and JSPS collaboration, the
 DGICYT contract number BFM2003-00856,
and the E.U. EURIDICE network contract no. HPRN-CT-2002-00311. 
This research is part of the EU
      Integrated Infrastructure Initiative
      Hadron Physics Project under contract number
      RII3-CT-2004-506078.

\vfill\eject
\end{document}